\documentclass[3p,times]{elsarticle}
\usepackage{amssymb}
\usepackage[english]{babel}
\usepackage{amsmath}
\usepackage{graphicx}
\DeclareMathOperator{\tr}{tr} 
\DeclareMathOperator{\diag}{diag} \DeclareMathOperator{\U}{U}
\usepackage[colorinlistoftodos]{todonotes}

\begin{document}

\begin{frontmatter}

\title{ Renormalization-group investigation of a superconducting $\U(r)$-phase transition using five loops calculations.}

\author[rvt]{G.A.~Kalagov}
\ead{KalagovG@gmail.com}
\author[rvt]{M.V.~Kompaniets}
\ead{mkompan@gmail.com}
\author[rvt]{M.Yu.~Nalimov\corref{cor1}}
\ead{Mikhail.Nalimov@pobox.spbu.ru}

\address[rvt]{ Department of Theoretical Physics, St. Petersburg State University,\\
Ulyanovskaya 1, St. Petersburg, Petrodvorets 198504, Russia}

\cortext[cor1]{Corresponding author}

\begin{abstract}
We have studied a Fermi system with attractive $\U(r)$-symmetric
interaction at the finite temperatures by the quantum field
renormalization group (RG) method. The RG functions have been
calculated in the framework of dimensional regularization and minimal subtraction scheme up to five loops. It has been found that for $r\geq 4$ the RG flux leaves the system's
stability region -- the system undergoes a first order  phase
transition. To estimate the temperature of the transition to
superconducting or superfluid phase the RG analysis for composite
operators has been performed using three-loops approximation. As the
result this analysis shows that for $3D$ systems estimated phase transition temperature is higher then well known theoretical estimations based on continuous phase
transition formalism.
\end{abstract}


\end{frontmatter}

\section*{Introduction.}

The investigation of quantum Fermi systems and the phase transitions
in these systems are the problems of permanent interest.
 To describe the quantum equilibrium Fermi system we  use  the  temperature
 Green functions formalism, quantum field theory methods and the
renormalization group approach.  The analysis is based on the
microscopic model with  local attractive interaction of the
``density-density''  type \cite{Abrikosov, Vasil'ev1, Kagan}. The model's
field action has the form
 \begin{equation}
\label{1} S_{\psi} =\psi^{\dag}_{\alpha}  (\partial_t-\frac
{\Delta}{2m}-\mu)\psi_{\alpha} -\frac{\lambda}{2}
\psi^{\dag}_{\alpha}\psi^{\dag}_{\gamma}\psi_{\gamma}\psi_{\alpha},
\end{equation} 
where  $\psi_{\alpha},\psi^{\dag}_{\alpha}$  describe the fermion fields at the finite temperature $T$, these fields are complex-conjugate elements of the Grassmann algebra and $\alpha = 1,\dots,r$, $r$ is the number of spin degrees of freedom, $\Delta$ is Laplace operator; $m$ is a mass of the particles; $\mu$ is the system's chemical potential; $\lambda = 4 \pi |a_s| /m$ is positive coupling constant and $a_s$ is the  scattering amplitude for interparticle $3D$-scattering; $t$ is the ``imaginary'' time and $t\in[0,\beta=1/T]$. All the necessary integrations and summations in formula \eqref{1} and similar expressions below are implied.
 It is also necessary to impose the antiperiodic boundary conditions with respect to ``imaginary'' time on the fermion fields.
\begin{equation}
\psi_{\alpha}({\bf p},0)=-\psi_{\alpha}({\bf p},\beta), \quad
\psi^{\dag}_{\alpha}({\bf p}, 0)=-\psi^{\dag}_{\alpha}({\bf
p},\beta).
\end{equation}

In the $r=2$ case, this   action (\ref{1}) corresponds to the
Bardeen-Cooper-Schrieffer theory and $\alpha =  \uparrow,\downarrow$
are the two possible spin projections. The theory describes low
temperature superconductivity in electron systems. We will consider
the case of arbitrary even values $r$. It can be corresponded to
systems of high spin fermions investigated recently \cite{atoms,
atoms1, atoms2, atoms3, atoms4, atoms5, atoms6, atoms7} or with the
electrons in solids which have a sublattice index and/or index
corresponding to a degenerations of zone structure \cite{grafen}.

Usually in the system under consideration the phase transition temperature is
determined by the appearance of an anomalous solution of the Dyson
equation \cite{Abrikosov}, and the order parameter of a
superconducting phase transition is given by means of the
composite operators $\left\langle \psi_{\alpha} \psi_{\gamma}
\right\rangle$ and $\left\langle \psi^{\dag}_{\alpha}
\psi^{\dag}_{\gamma} \right\rangle$. To investigate this model using renormalization group method the action is transformed by introducing the new boson fields $\chi, \chi^{\dag}$
\cite{Nalimov, Kalagov}. The action of the form
\begin{equation}
\label{2}
S_{\psi,\chi}=\psi^{\dag}_{\alpha}(\partial_t+\varepsilon_{\bf p})\psi_{\alpha} +\frac{1}{2\lambda}\tr \chi \chi^{\dag} -\frac{1}{2}
\psi^{\dag}_{\alpha}  \chi_{\alpha \gamma}
\psi^{\dag}_{\gamma}-\frac{1}{2}\psi_{\alpha} \chi_{\alpha \gamma}
^{\dag}\psi_{\gamma}
\end{equation}
was considered, where $\varepsilon _{\bf p}= {\bf p}^2/(2m)-\mu$.

It can be easily proven that the integration $\exp (-S_{\psi,
\chi})$ over the fields $\chi, \chi^{\dag}$ leads to $\exp
(-S_{\psi})$. The new fields are complex skew-symmetric matrices
because the fields $\psi$, $\psi ^{\dag}$ are Grassmann variables.
The Schwinger equations
\begin{eqnarray}
\nonumber
\langle \chi^{\dag}_{\alpha \gamma}+\lambda \psi^{\dag}_{\alpha}\psi^{\dag}_{\gamma}\rangle =0,\\
\nonumber \langle \chi_{\gamma \alpha}-\lambda
\psi_{\alpha}\psi_{\gamma} \rangle =0.
\end{eqnarray}
show that the $\chi, \chi^{\dag}$ determine the order parameter of a
phase transition.

The integration of $\exp (-S_{\psi, \chi})$ over the fermion fields
$\psi$, $\psi ^{\dag}$ leads to the new action for the boson fields
$\chi, \chi^{\dag}$
\begin{equation}
S_{\chi} =\frac{1}{2\lambda} \tr \chi \chi^{\dag}-\tr \ln
\begin{pmatrix}
   - \chi^{\dag} &  -i \omega_s-\frac {\Delta}{2m}-\mu \\
    -i \omega_s +\frac {\Delta}{2m}+\mu &- \chi
  \end{pmatrix},
\end{equation}
here $\omega_s = \pi T (2 s +1)$ are Matsubara frequencies, $s\in \mathbb{Z}$. Using the Taylor
expansions for $\ln(1+\dots)$ we can rewrite the action as
\begin{eqnarray}
\label{3}
 S_{\chi} =\frac{1}{2\lambda} \tr \chi\chi^{\dag} + \frac{1}{2}\parbox[b][0.5cm][t]{3.2cm}{\includegraphics[width=90pt]{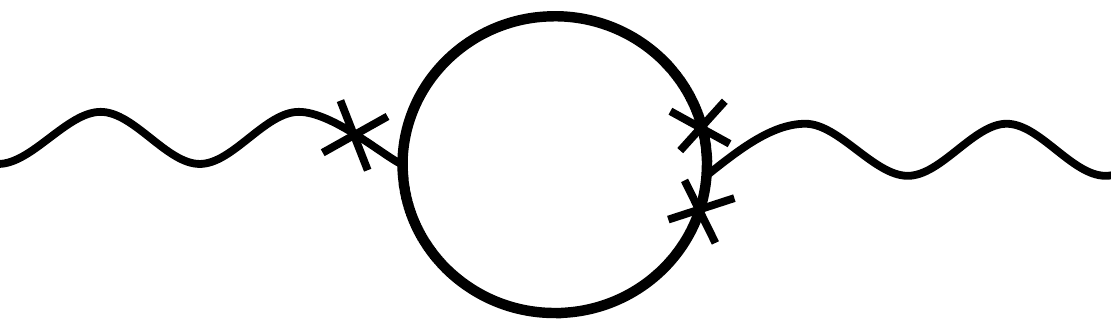}}+ \frac{1}{4} \parbox[b][0.6cm][t]{3.2cm}{\includegraphics[width=90pt]{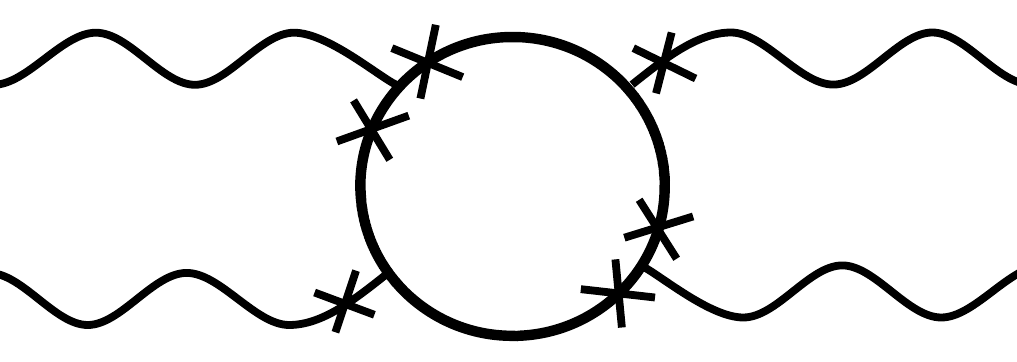}}+ \dots ,
\end{eqnarray}

here wave lines denote the field $\chi $, $\chi ^{\dag}$, the
plain lines denote free $ \left\langle \psi \psi^{\dag}
\right\rangle $ propagators, cross corresponds to $\psi ^{\dag}$,
$\chi ^{\dag}$ fields.

To obtain the effective action in the infra red (IR) region we have
to present (\ref{3}) in the form of a Ginzburg-Landau functional by
expanding all diagrams in the external momenta ${\bf p}$ and
 frequencies. Then $\chi ^{\dag}$, $\chi $ fields can be considered as $t$-independent. As a result the effective action has the form
\begin{eqnarray}
\label{act}
 S_{\chi} =\tr\chi^{\dag}({c_0}{\bf p}^2 +\widetilde{\tau}_0)\chi+\frac{\widetilde{g}_{01} }{4}\tr (\chi\chi^{\dag})\tr (\chi\chi^{\dag}) +\frac{\widetilde{g}_{02} }{4}\tr (\chi\chi^{\dag}\chi\chi^{\dag}).
\end{eqnarray}
The term with $\widetilde{g}_{01}$ coupling constant was included to obtain the
multiplicatively renormalized theory. The parameters of the action
$c_0$ and $\widetilde{g}_{02}$ are positive and they can be
calculated from the expressions
\begin{equation}
\label{g} \widetilde{g}_{01} = 0, \quad \widetilde{g}_{02} = \beta
T\sum\limits_{\omega_s}\int \frac{d^D{\bf
k}}{(2\pi)^D}\frac{1}{(\omega_s^2+\varepsilon_{\bf k}^2)^2}, \quad
 \widetilde{\tau}_0 = \frac{\beta}{2\lambda}-\frac{\beta}{2}T\sum\limits_{\omega_s}\int \frac{d^D{\bf
k}}{(2\pi)^D}\frac{1}{\omega_s^2+\varepsilon_{\bf k}^2},
\end{equation}

\begin{equation}
\label{c0} c_0 = \left.-\frac{\beta}{2}T \partial^2_{\bf p}\sum\limits_{\omega_s}\int
\frac{d^D{\bf k}}{(2\pi)^D}\frac{1}{(i\omega_s+\varepsilon_{\bf k})(-i\omega_s+\varepsilon_{\bf k+p})}\right|_{{\bf p}=0},
\end{equation}
here  $D$ is dimension of space, $\omega_s$ are Matsubara frequencies. The integration over ${\bf k}$ is performed in a narrow neighborhood of the Fermi surface $|\varepsilon_{\bf{k}}-\mu|<\delta$. The parameter $\delta $ can be
similar to Debye frequency $\omega_D$ for the system of electrons in
the solids or similar to Fermi energy $\varepsilon _F$ for ultra
cold atoms systems and $\delta = (2/e)^{7/3}\varepsilon_F \approx 0.49 \varepsilon_F$ for such systems \cite{Gorkov}.

The IR behavior of the model (\ref{act}) was studied in \cite{Nalimov,
Kalagov}. The renormalization group (RG) investigation  in the
framework of $\varepsilon=4-D$ expansion in one-loop approximation
\cite{Nalimov} and then in three-loop approximation \cite{Kalagov}
establishes the absence of IR-stable fixed points for even values of
$r\ge 4 $. It was found that the stability criterion for action
(\ref{act})  (the condition for positive definiteness of
an interaction) can be formulated as the inequality
\begin{equation}
\label{stab} {g}_2 + r {g}_1>0,
\end{equation}
for ${g}_2>0$.

Moreover solutions of the RG equations for the invariant charges in
one-loop approximation \cite{Nalimov} show that the system loses the
stability before the continuous phase transition occurs. It was
supposed that a first-order phase
transition takes place here,/ and this phase transition can be considered as one of the possible reasons of high temperature superconductivity.

Then the similar behavior was
confirmed in \cite{Kalagov} in the
three-loop RG analysis of the 3D  and 2D models. But it was found
that the three-loop approximation is not sufficient to ensure
an accurate calculation of the phase transition temperature.
Therefore we have to develop our analysis up to five-loop calculations, that is
the maximal order available now in the framework of $\varepsilon$-
expansion\cite{phi45loop}.

In Sec.2 we describe the five-loop RG analysis  of the model
investigated with $r \geq 4$. According \cite{Nalimov} there is no IR stable fixed point in the framework of $\varepsilon$ expansion.
Thus instead of seeking fixed points of the RG equation we restrict
ourselves to  analysis of phase trajectories. It was indicated in
\cite{Nalimov} that the equations for the invariant charges can be
constructed in the $\varepsilon$-expansion form.

To analyze the phase portrait of these equations in the physical
space dimensions ($\varepsilon =1$, $\varepsilon =2$) we must resum
the terms calculated, for instance, using the Borel resummation
technique. Such a resummation requires knowing the higher orders
asymptotics (HOA) of the $\varepsilon $ expansion. The HOA of the considered model
was determined in \cite{Kalagov} using
methods of instanton analysis \cite{Lipatov}. The analysis and the
results obtained are described briefly in Sec. 3. It is interesting
to note that we  have found several instantons with different matrix
structures. These instantons are essential in a Borel resumming  at different values of the charges $g_1$, $g_2$.

In Sec.4 we resum and solve numerically the RG equations for
invariant charges. It is
confirmed that the invariant charges in the $3D$ model cross the boundary (\ref{stab}) of the stability domain of the action (\ref{act}).
As for the $2D$ model, it is found that
five-loop approximation is not sufficient yet for the accurate
description of the phase transition type. Our results show that the
type  depends on the initial value of the coupling constant
$g_{20}$.

In Sec.5 the first order phase transition is studied in $3D$ and
$2D$ model to find the real phase transition temperature. The
additional terms ($\sim \chi ^6$) were introduced in action
(\ref{act}). These terms are IR irrelevant for critical behavior,
but are relevant for the first order phase transition description.
They are renormalized as a composite operators in three-loop
approximation. Their contributions to the state equation are Borel
resummed and the phase transition temperatures estimated.

\section{Renormalization group analysis}

 The renormalized action of the considered model is given by the expressions
\cite{Nalimov}
\begin{equation}
\label{actrg} S_R  =Z_{\chi}^2
\tr\chi^{\dag}(-\Delta)\chi+Z_{\tau}\,Z_{\chi}^2\, \tau
\tr\chi^{\dag}\chi + Z_{g_1}\,Z_{\chi}^4\,
M^{\varepsilon}\frac{g_{1}}{4} \left(\tr \chi\chi^{\dag}\right)^2 +
Z_{g_2}\,Z_{\chi}^4\,M^{\varepsilon}\frac{g_{2}}{4}\tr
\chi\chi^{\dag}\chi\chi^{\dag}.
\end{equation}
This expression is obtained by the multiplicative renormalization
 \begin{equation}
 \label{bar}
 g_{0j} \rightarrow g_{j}M^{\varepsilon}Z_{g_i}, \quad \chi \rightarrow \chi Z_{\chi}, \quad \tau_0 \rightarrow \tau Z_{\tau},
\end{equation}
here the parameter $M$ is a so-called renormalization mass; $g_{1},
g_{2}$ are dimensionless renormalized  coupling constants, index
zero denotes bare parameters. In this paper we use the dimensional regularization,
$\varepsilon$-expansion and the minimal subtraction  scheme
(MS-scheme) \cite{Vasil'ev}. The bare parameters
$g_{0j}$ and $\tau_0$ are associated with the microscopic parameters
(\ref{c0}, \ref{g}) by the relations $g_{0j} =
\widetilde{g}_{0j}/ c_0^2$, $\tau_{0} = \widetilde{\tau}_{0}/c_0$ and
$\chi \rightarrow \chi/\sqrt{c_0}$.

 \begin{figure}
 \label{vert}
\includegraphics[width=12cm]{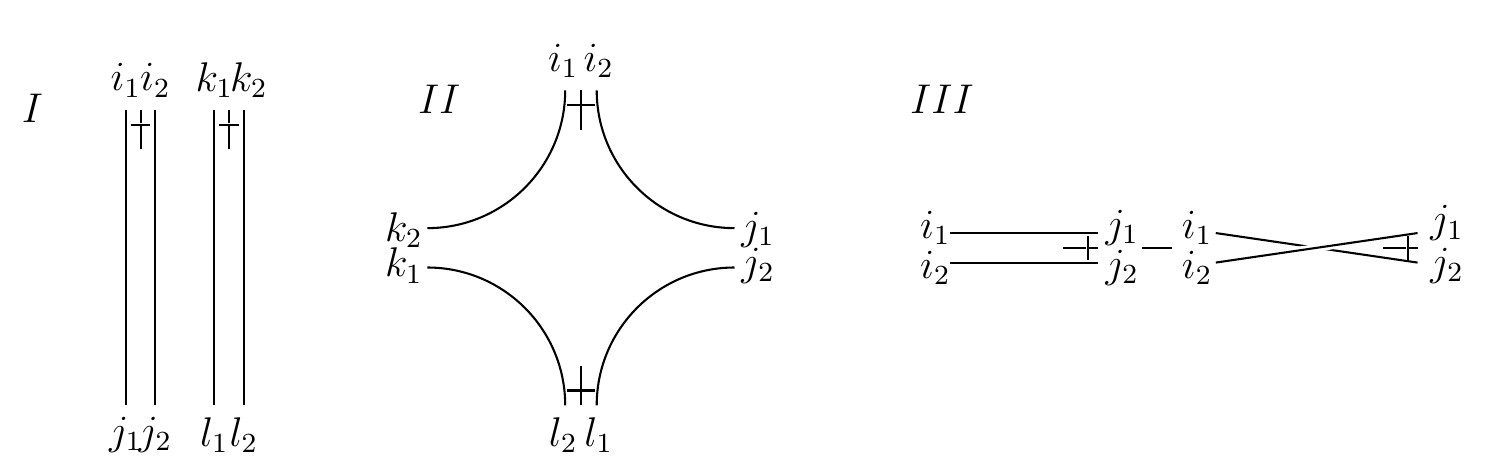}
\caption{Tensor structures of the vertices and propagator: vertex I
corresponds to $g_1$, vertex II corresponds to $g_2$, and vertex III
corresponds to the propagator.}
 \end{figure}

Let us introduce basic elements of the Feynman diagrammatic
techniques for the model. In momentum representation the free
propagator  has the form
\begin{equation}
\notag \Pi_{i_1i_2}^{j_1j_2} = \frac{W_{i_1i_2}^{j_1j_2}}{{\bf
k}^2+\tau_0} \quad \text{and} \quad W_{i_1i_2}^{j_1j_2} \equiv
\frac{1}{2}(\delta_{i_1j_1} \delta_{i_2j_2}-\delta_{i_1j_2}
\delta_{i_2j_1}),
\end{equation}
here $\delta_{ij}$ is Kronecker symbol; $\bf{k}$ is the momentum or
the wave vector ($\hbar=1$). The tensor $W_{i_1i_2}^{j_1j_2}$ is
antisymmetric with respect to the transpositions of its indexes $i_1
\leftrightarrow i_2$ and $j_1 \leftrightarrow j_2$ , and symmetric
with respect to the transposition of the pairs $(i_1, i_2)
\leftrightarrow (j_1, j_2)$. One can write the tensor structures for
the vertices $g_1$ and $g_2$ too, but we give only their graphical representation (Fig.\ref{vert}) , the vertices antisymmetrization  is
implied.

  In MS-scheme all renormalization constants have the form of the poles in
  $\varepsilon$
\begin{equation}
\notag Z_e = 1 + \frac{Z^1_e}{\varepsilon}+O\left(\frac 1
{\varepsilon ^2}\right),\quad e=(g_j, \tau, \chi)
\end{equation}
here $ Z^1_e(g_1,g_2)$  denotes the residue at the simple pole in $\varepsilon$ for the corresponding renormalization constant.
Let us remark that the interaction
$\left(\tr \chi\chi^{\dag}\right)^2$ must be included for the
multiplicative renormalizability of the theory. It is easy to verify
that the corresponding counterterms appear due to the
renormalization of the theory starting with the simplest one-loop
diagram.

The  RG-functions (the coefficients of the RG equation
\cite{Vasil'ev}) are defined by the relations
\begin{equation}
\label{rgf} \beta_{g_j} = \widetilde{D}_M Z_{g_i}, \quad \gamma_{e}
=\widetilde{D}_M \ln Z_ {e},
\end{equation}
here $\widetilde{D}_M$ is the differential operator $M\partial_M$ at
fixed bare parameters, $\beta_{g_j}$ are  beta-functions of charges
$g_i$, and the functions $\gamma_{e}$ are  anomalous dimensions for
parameters $e$. In the MS-scheme the RG-functions are connected with
the renormalization constants by the following expressions
\cite{Vasil'ev}
\begin{equation}
\label{rg} \beta_{g_j} = -g_j(\varepsilon+\gamma_{g_j}), \quad
\gamma_{e} = -g_k\partial_k Z^1_{e}.
\end{equation}
Program ``FORM'' was used\cite{Varm} for  the tensor structure calculations of the graphs. Tensor structure of the graph can be factorized, and the rest of the diagram is equivalent to diagrams of the scalar $\Phi^4$ model, the values for these diagrams are taken from well known five loop calculations of the $O(n)$-symmetric $\Phi^4$  model \cite{phi45loop,phi45loopnum}.   Finally, in the five-loop approximation (about
120000 diagrams), the RG-functions of
 the theory were calculated. The rescale of the charges  $g_i
\rightarrow g_i/{16\pi^2}$ was used. Results of our calculations
were controlled for $r=2$, $r=3$. In these cases the model
(\ref{actrg}) is equivalent to the $O(2)$- and $O(6)$-$\Phi^4$
models with vector order parameter, respectively.

The  RG equation leads to the known equations  for the invariant
coupling constants
\begin{equation}
\label{RGi}
\partial_{\xi} \bar{g}_i  =\frac {\beta_{{g}_i}}{2+\gamma_{\tau}}, \quad \left.\bar{g}_i\right|_{\xi=0}= g_i, \quad \text{where} \quad \xi\equiv\ln\frac{\tau}{M^2}.
\end{equation}
The infra red (IR) regime $\xi \rightarrow -\infty$ is
usually connected with fixed points ($g^*_1,g^*_2$) that are determined by the
conditions $\beta_{g_i}(g^*_1, g^*_2)=0$ for all indices $i$. The
fixed point is IR-stable, if the matrix  $\omega_{ij} \equiv
\partial_{g_j} \beta_{g_i}(g^*_1, g^*_2)$ is  positively
defined. However, it was found, in the one-loop approximation
\cite{Nalimov}, that
 these points do not exist for $r \geq 4$. There is the IR-stable fixed point in the
 model at $r=2$,  this point describes the critical behavior of the
 superconducting phase transition in systems with $1/2$-spin fermions.

We will not search for the possible fixed point in the five-loops
approximations of the model considered, instead of this trajectories of the
invariant charges will be studied in the next sections.

\section{Instanton analysis}

Consider the equations (\ref{RGi}) with $\beta $-functions
(\ref{b1},\ref{b2})(see Appendix). After the  scaling of the charges $\bar{g}_i$
and the dynamical variable of the RG equation $\xi$ as $\bar{g}_i
\rightarrow \varepsilon \bar{g}_i$ and $\xi \rightarrow  \xi/
\varepsilon$, we get eq. (\ref{RGi}) in the  form
\begin{equation}
\begin{split}
\label{g1}
&\partial_{\xi} {\bar g}_i = -{\bar g_i}+\sum\limits_{N=0}^{K} \varepsilon^N B^{(N)}_{i}({\bar g}_1,{\bar g}_2), \quad i=1,2\\
&\left.{\bar g}_i\right|_{\xi=0} = g_i.
\end{split}
\end{equation}
Explicit expressions of the $B^{(N)}_{i}({\bar g}_1,{\bar g}_2)$ can
be obtained from (\ref{b1}, \ref{b2}). In our case $K=4 $ (five-loop
approximation). Equations (\ref{g1}) can be solved in the
form of $\varepsilon$-expansion with the formally small parameter
$\varepsilon$. Similar to \cite{Nalimov} we will consider numerical
solution of the equations (\ref{g1}). As usual, the
$\varepsilon$-expansion in the right hand side of equations (\ref{g1})
is asymptotic expansion with zero radius of convergence. Then the
equations (\ref{g1}) must be resummed to obtain results at
physical points $\varepsilon =1$ or $\varepsilon=2$. The resummation
process requires knowledge about the asymptotic behavior
$B^{(N)}_{i}({\bar g}_1,{\bar g}_2)$ at $N\rightarrow \infty$. Such
asymptotic behavior is called a higher-order asymptotic (HOA) and
was investigated in \cite{Kalagov}   in the model considered.

Let us recall the main details of the analysis \cite{Kalagov}. The
investigation of the asymptotic behavior of higher-order
perturbation corrections proposed in \cite{Lipatov} is based on the
saddle-point expansion of the path integral (instanton approach).
Calculation method for the HOA for renormalization constants in MS
scheme developed in \cite{Komarova} was used. Partially renormalized
Green function was considered where subtractions of all the
divergences in subgraphs up to order $N-1$ are supposed. The
coefficients $G_{2k}^{(N)}$ of the expansion in the parameter
$\varepsilon$ of the $2k$-point Green function
\begin{eqnarray}
\label{gren} G_{2k}(\varepsilon,x_1,\dots, x_{2k}) =
W^{-1}\int\mathcal{D}\chi \mathcal{D}\chi^{\dag}
\chi(x_1)\chi(x_2)^{\dag} \dots \chi(x_{2k-1})\chi(x_{2k})^{\dag}
e^{-S_R},
\end{eqnarray}
\begin{equation}
\notag W = \int\mathcal{D}\chi \mathcal{D}\chi^{\dag} e^{-S_R}.
\end{equation}
can be calculated in high orders ($N\rightarrow\infty$) with the use
of the saddle-point method in the integral representation
\cite{Lipatov}
\begin{equation}
\label{koshi} G_{2k}^{(N)}(x_1,\dots, x_{2k}) = \frac{(-1)^N}{2\pi
i}\oint\limits_{\gamma} \frac{d \varepsilon
G_{2k}(\varepsilon,x_1,\dots, x_{2k})}{(-\varepsilon)^{N+1}},
\end{equation}
where $\gamma$ is a closed contour encircling the origin in the
complex plane of $\varepsilon$. As usual \cite{Lipatov}, we will
find the HOA at $\tau =0$ and $D=4$. After the rescaling of the
parameters $g_i \rightarrow g_i/N$, $\chi \rightarrow \sqrt{N}\chi$,
$\chi^{\dag} \rightarrow \sqrt{N}  \chi^{\dag}$ the variational
equations for functional  $S_R+\ln(-\varepsilon)$ with respect to
the field variables and $\varepsilon$ takes  the form
\begin{equation}
\begin{split}
\label{var1}
&-\Delta \chi +\frac{ \varepsilon g_1}{2} \chi \tr \chi\chi^{\dag}+\frac{\varepsilon g_2}{2}\chi\chi^{\dag}\chi  =0,\\
&-\Delta\chi^{\dag} +\frac{\varepsilon g_1}{2} \chi^{\dag} \tr
\chi\chi^{\dag}
+\frac{\varepsilon g_2}{2}\chi^{\dag}\chi\chi^{\dag} =0,\\
&\int d{\bf x}\left\{\frac{\varepsilon g_1}{4} \left(\tr
\chi\chi^{\dag} \right)^2+\frac{\varepsilon g_2}{4} \tr
\chi\chi^{\dag}\chi\chi^{\dag}\right\}= -1.
\end{split}
\end{equation}
Similar to \cite{Lipatov}, the counterterms $Z_{e}-1$ in action
(\ref{actrg}) are irrelevant for the calculation of the stationary
points. For matrix fields $\chi$, $\chi^{\dag}$ we can assume
without loss of generality the block-diagonal  Phaff's form
containing of $p=r/2$ blocks
\begin{equation}
\label{pff} \chi = \diag(s_1 \sigma,\dots, s_p\sigma), \quad
\chi^{\dag} = -\diag(s^*_1 \sigma,\dots, s^*_p\sigma), \quad
\sigma=\bordermatrix{
   &            \cr
  & 0  & -1   \cr
  & 1  & 0    \cr },
\end{equation}
with some complex functions $s_j({\bf x} )$. Any skew-symmetric
matrix can be reduced to this form by some unitary transformations
$U(r)$. The equations
 (\ref{var1}) and  (\ref{pff}) yield the system of equations for $s_j({\bf x} )$
\begin{equation}
\label{ds} -\Delta s_i({\bf x})  + \varepsilon g_1
\sum_{k=1}^{p}|s_k({\bf x}) |^2s_i({\bf x})  + \frac{\varepsilon
g_2}{2} |s_i({\bf x}) |^2 s_i({\bf x}) =0.
\end{equation}
We  seek $s_j({\bf x} )$ in the  form
\begin{equation}
\label{s}
 s_i({\bf x})=\frac{\alpha_i \; y^{-1}}{|{\bf x}-{\bf x}_0|^2 + y^2}, \quad
\alpha_i \in \mathbb{C},
\end{equation}
similar to solutions of variational equation for the scalar $\Phi^4$
model. Functions $s_i({\bf x})$ depend on ${\bf x}_0$ and  $y$ --
arbitrary parameters reflecting the translational and dilatation
invariance of the theory. Then the Faddeev-Popov method was used
similar to \cite{Lipatov}. The method fixes the values of the free
parameters for each realization of $s_i({\bf x})$. Substituting
(\ref{s}) in (\ref{ds}), we get the system of algebraic equations for
constants $\alpha_i$
\begin{equation}
\label{a}
 8 \; \alpha_i +  \varepsilon  g_1 \sum_{k=1}^{p}|\alpha_k|^2 \alpha_i +\frac{ \varepsilon g_2}{2} |\alpha_i|^2 \alpha_i =0.
\end{equation}
One can see that the stationary solutions may contain  $m =0,\dots,p-1$
zero blocks  with $|\alpha_i|=0$ and $n =p,\dots,1$ blocks with
$|\alpha_i|^2 = -16/(2 n \varepsilon g_1 + \varepsilon  g_2)$, and
$n+m=p$. Phases of the complex numbers $\alpha_i$ are not fixed, they
are arbitrary parameters as well as ${\bf x}_0$ and $y$. In addition
to parameters $y$,  ${\bf x}_0$ and phase factors,  there is also an
invariance under unitary transformations $U(r)$. Thus the number of
zero modes is determined by $r^2-2r +n +5$. This number  influences
the exponent of $N$ in the HOA.

Combining the instanton solutions (\ref{s}), the Pfaff's form
(\ref{pff}) and  the third equation (\ref{var1}), we get stationary
point in $\varepsilon$ parameter as
\begin{equation}
\label{z} \varepsilon_{st}(n) = -\frac{4n}{3} \frac{1}{ 2n g_1+g_2}.
\end{equation}

Similar to  \cite{Komarova}, the beta-functions HOA can be obtained
from the HOA for Green functions  residue at the simple pole in
$\varepsilon$.
\begin{equation}
\label{bn} \beta_i^{(N)}(g_1, g_2) =  const_i N! N^{b_n} \left( -a
\right)^N\left(1+O\left( N^{-1}\right)\right) ,
\end{equation}
here $const_i$ -- some constants not essential for future analysis,
$b_n = (r^2 - 2 r + n+ 11 )/2$
 and  $a=\underset{n}{\max} |a(n)|$, $a(n) = -1/\varepsilon_{st}(n)$.
One can see from  (\ref{z}), that $a(n)$  depends on values of $g_i$, therefore
the largest of all $a(n)$ gives the largest contribution to the HOA.
Thus the perturbation series in the
parameter $\varepsilon$ have zero radius of convergence in the
theory with the action (\ref{actrg}). For this reason, it is
necessary to use some procedures of resummation e.g. the Borel
method.

\section{Solution of the RG equations. }
\subsection{Resummation of the RG equations.}

 Let us recall the basic expressions for the Borel resummation \cite{hagen}. We assume that there is a function $Q(\varepsilon)$ defined as a series on the parameter $\varepsilon$
\begin{equation}
\label{physquan}
  Q(\varepsilon)=\sum_{N\ge0} \varepsilon^N Q^{(N)} ,
\end{equation}
and the higher-order asymptotics of the series coefficients are
determined by expression (\ref{bn}). The Borel transform of the
series (\ref{physquan}) is given by the relations
\begin{equation}
\label{bor} Q(\varepsilon) = \int\limits_{0}^{\infty} dt ~ e^{-t}
t^{b_0} B(\varepsilon t), \quad B(t)=\sum_{N\ge0} B^{(N)} t^N, \quad
B^{(N)} = \frac{Q^{(N)}}{\Gamma (N+b_0+1)},
\end{equation}
where $b_0$ is an arbitrary parameter. The known asymptotic
expansion (\ref{bn}) together with several assumptions about the
analytic properties of $B(t)$  allow one to resum series
(\ref{physquan}) using (\ref{bor}) and to obtain a more precise
value of $Q(\varepsilon)$. According to (\ref{bn}), the series
$B(t)$ given by (\ref{bor}) converges in the circle $|t| < 1/a$,
because $B^{(N)} \sim \left( -a \right)^N  N^{b_n-b_0}$ as
$N\rightarrow\infty$. The nearest singularity of the series is
located on the negative real half-axis at the point $t = -1/a$. Then
the integration contour over $t \in [0, +\infty)$ intersects the
boundary of the circle of convergence for expression (25) at the
point $1/a$. The problem of analytical continuation of (\ref{bor})
beyond the convergence domain $|t| < 1/a$ can be solved either by
the method of the conformal mapping of the complex plane or by
Pad\'e approximation method \cite{hagen}. Below we will use the
conformal mapping method, because it is controlled by HOA.
Furthermore, it leads to more accurate results than other methods
(see \cite{sergeev}).  In our case the position of the
$B(t)$-function poles depends on the position of the invariant
coupling constants in the ($\bar{g}_1$,  $\bar{g}_2$) plane.

For example, let us consider a system (\ref{RGi}) with $r=4$. There
are two kinds of instantons in the model. For instanton containing
one non-zero block we get $a(1) = 3 (2 g_1+g_2)/4$. Otherwise,
instanton has two non-zero blocks and $a(2) = 3 (4 g_1+g_2)/8$. Therefore
in stability sector \eqref{stab} there are two regions in the plane ($\bar{g}_1$, $\bar{g}_2$)
where series for $B(t)$  have different analytical properties:

Region $I$: if the invariant coupling constants satisfy the
condition $8\bar{g}_1+3\bar{g}_2 > 0$, then  $|a(1)|>|a(2)|$. The
nearest singularity of the $B(t)$ is $t=-1/a(1)$;

Region $II$: if the invariant coupling constants satisfy the
condition  $8\bar{g}_1+3\bar{g}_2 \leq 0$, then $|a(2)|>|a(1)|$.  In
this case the nearest singularity of the $B(t)$ is located on the
positive real half-axis at the point $t=1/|a(2)|$.

Thus, the plane ($\bar{g}_1$,  $\bar{g}_2$) is divided by the line
$8 \bar{g}_1 + 3 \bar{g}_2 =0$.  Above this boundary the analytical
properties of $B(t)$-functions are determined by one non-zero block
instanton, under the  boundary only the two non-zero blocks
instanton influences the properties of function $B(t)$.

 The initial values of the invariant coupling constants  are located in the region $I$. Let us apply conformal mapping method for the invariant coupling constants located in this region. Usually the conformal map of the complex plane is chosen in the form \cite{Zinn, hagen}
\begin{equation}
\label{cm} u(\varepsilon)=\frac{\sqrt{1+a \varepsilon}-1}{\sqrt{1+a
\varepsilon}+1} \quad \Leftrightarrow\quad \varepsilon(u)=\frac{4
u}{a (u-1)^2}.
\end{equation}
The series (\ref{physquan}) can be rewritten in terms of the
variable $u$ as
\begin{equation}
\begin{split}
\label{rrborel}
&B(\varepsilon)=\sum_{N\ge0} B^{(N)} \varepsilon^N =\sum_{N\ge0} U^{(N)} u^N,\\
& U^{(0)}=B^{(0)},\quad U^{(N)}=\sum^N_{m=1} B^{(m)} (4/a)^m
C^{N-m}_{N+m-1}, \quad N \ge1,
\end{split}
\end{equation}
then the conformal Borel map of the quantity $Q$ looks as follows
\begin{equation}
\label{cm_integ} Q(\varepsilon) =\sum_{N\ge0} U^{(N)}
\int\limits_{0}^{\infty} dt ~ t^{b_0} e^{-t} u (\varepsilon t) ^N.
\end{equation}
Usually, the parameter $b_0$ is chosen to weaken the
singularity of the Borel transform (\ref{bor}) at the point $t =
-1/a$. It is fixed by the relation $ b_0 = b_n+3/2$ \cite{Zinn,
hagen}.

In the region $II$ the singularity  of the function $B(t)$ is located on
the  positive real half-axis, thus the conformal mapping
method can not be used.

\subsection{Numerical analysis of the RG-equations.}

Combining the RG equations (\ref{g1})  and resummation formula
(\ref{cm_integ}), we have resummed RG equations for the
invariant coupling constants
\begin{equation}
\begin{split}
\label{g1p}
&\partial_{\xi} {\bar g}_i = -{\bar g}_i+\sum\limits_{N=0}^{K} U_i^{(N)} \int\limits_{0}^{\infty} dt~ t^{b_0} e^{-t} u (\varepsilon t) ^N ,\\
&\left.{\bar g}_i\right|_{\xi=0} = g_i.
\end{split}
\end{equation}
Note that $U_i^{(N)}$ and $u(t)$ are functions of the variables
${\bar g}_i$. This system of equations (\ref{g1p}) can be solved by
standard finite-difference method.

\begin{figure}
\centering
\includegraphics[width=7cm]{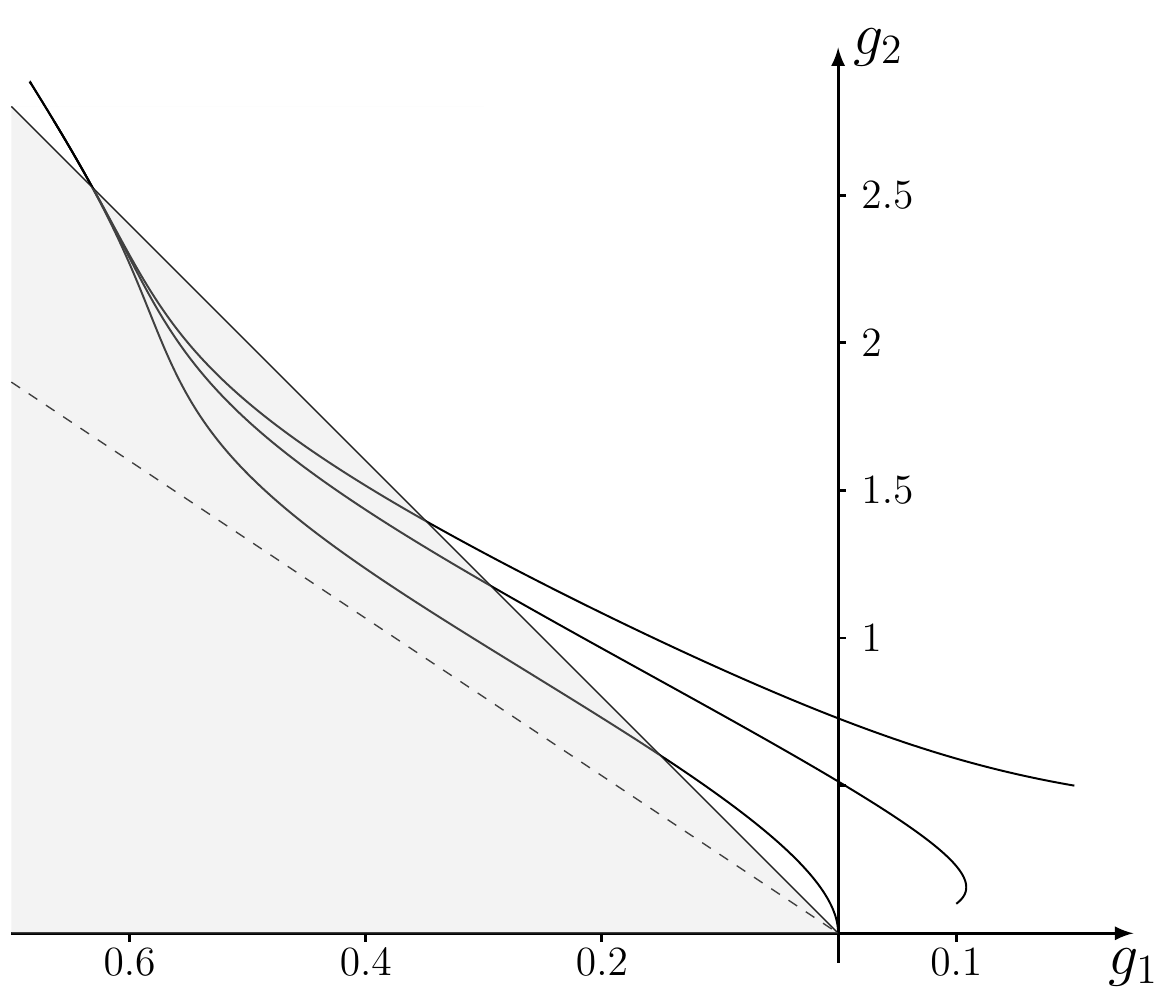}
\caption{ Trajectories of the running coupling constants at $D=3$ and
$r=4$;  dashed line -- the boundary of applicability of the
resummation method.} \label{ris3}
\end{figure}
\begin{figure}
\centering
\includegraphics[width=7cm]{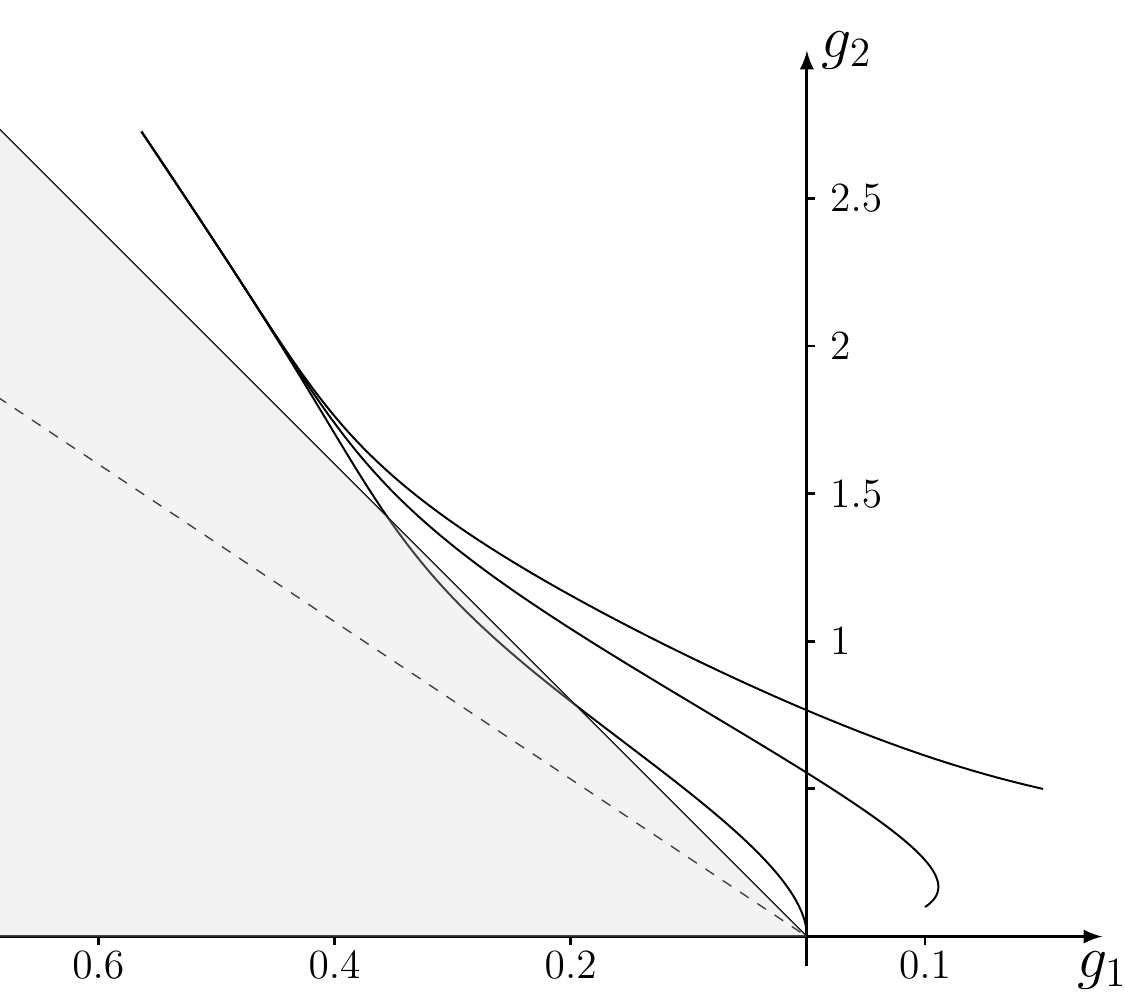}
\caption{Trajectories of the running coupling constants at $D=2$ and
$r=4$;  dashed line -- the boundary of applicability of the
resummation method.} \label{ris4}
\end{figure}

\begin{figure}
\centering
\includegraphics[width=7cm]{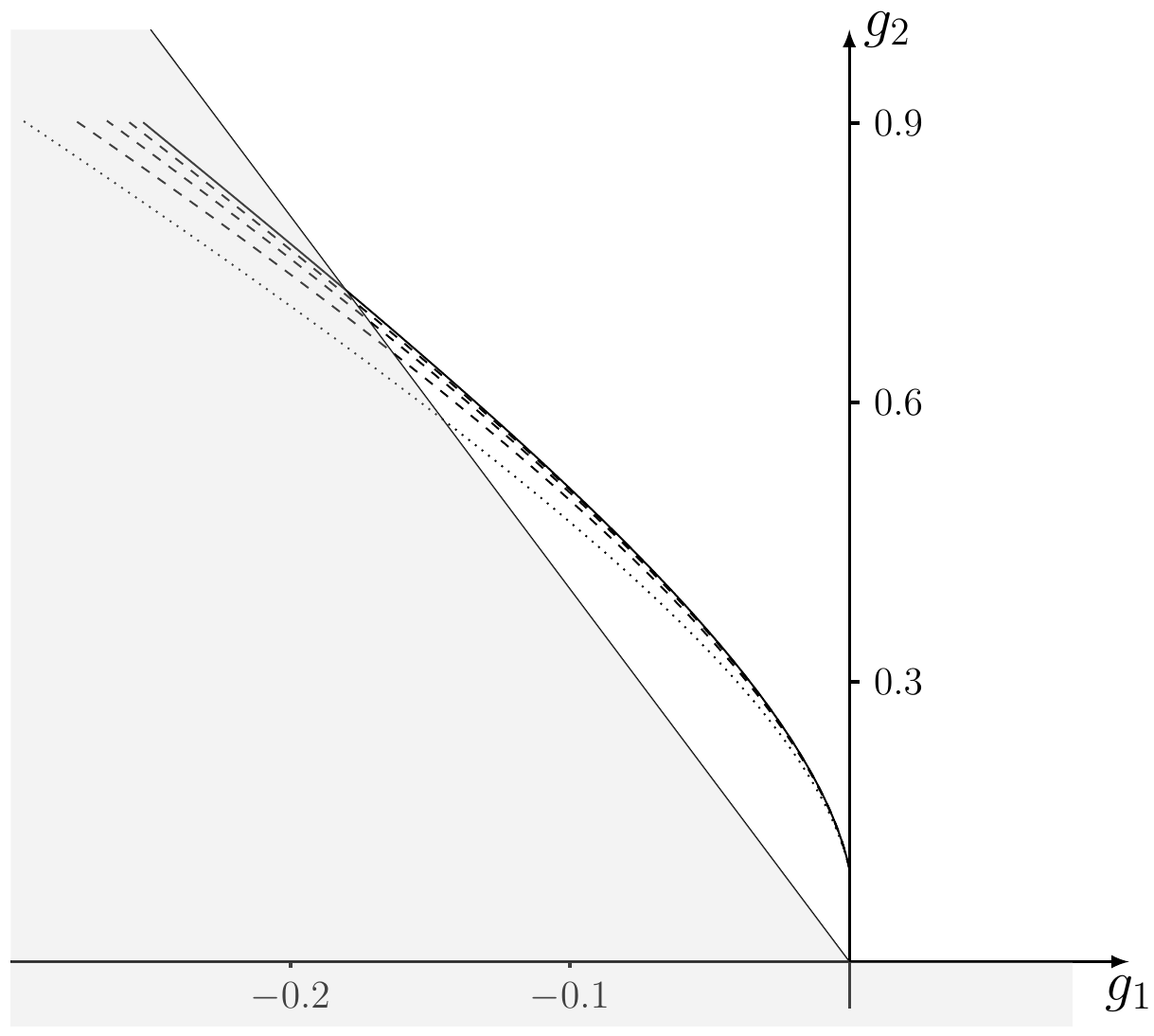}
\caption{The solutions of the RG equations($D=3, r=4$) at different numbers of
calculated loops: 1-loop -- dotted line, 5-loops -- solid line.}
\label{loops}
\end{figure}

The results of numerical solutions of the system
(\ref{g1p}) for $r=4$ are shown as an example in  Fig.\ref{ris3} at
$\varepsilon=1$ and Fig.\ref{ris4} at $\varepsilon=2$. Fig.
\ref{ris3} shows that in three dimensional model the invariant
charges trajectories starting with different initial values cross
the boundary of the action stability domain at some value $\xi_0$ of
the parameter $\xi$. Similar behavior is observed for different
values $r \geq 4$.

Fig.\ref{loops} shows how the trajectories of invariant charges
depend on the order of loops calculations for $D=3$. We can state that
five-loops approximation is sufficient to ensure the loss of the
action stability and accurate calculation of  $\xi_0$. Moreover,
numerical analysis shows that solutions of the Cauchy  problem
(\ref{g1p}) are stable under small perturbations of initial
conditions.

It is interesting to note, that we have found the IR stable fixed
point of RG equation. According  to \cite{Nalimov,Kalagov} there is no
IR stable fixed point for $\beta $-functions in one, two and
three-loops approximations. The fixed point appear in four-loops.
But the five-loops corrections essentially change  the
 position of this fixed point, so we can guarantee neither existence nor
position of this fixed point.

In $D=2$ ($\varepsilon = 2$) case the IR stable fixed point is found
in four- and five- loop approximation too. But in difference to three-loops
ones \cite{Kalagov} only rare
trajectories of the invariant charges cross the line of the action
 stability (\ref{stab})  according to Fig.(\ref{ris4}). These trajectories are connected with very small initial
values of renormalized coupling constants.

Our calculations are valid if the invariant coupling constants are
in the region $I$. In the region  $II$ the series can not be
resummed by the Borel method. Finally note that $a(1)$ gives the
largest contribution to the HOA in the stability domain and  in a
neighborhood of the stability boundary for any $r>2$. One can assume
that the phase transition occurs near the boundary of stability. For
this reason resummation process can be made only for $a(1)$.

\section{The phase transition description.}
The loss of the action stability is usually considered as a mark of
the first-order phase transition. But obviously, it is not possible to
claim that $\xi_0$ defines the first-order phase transition
temperature; only metastable states appear in the system at
$\xi=\xi_0$. To answer the question when the new state of the system
(with a condensate) in fact becomes stable, i.e., to determine the
phase transition temperature, more accurate analysis is necessary.

Because the interaction terms ($\sim \chi ^4$) of the action
(\ref{actrg}) are not positively defined now, we have to take into
account the next term ($\sim \chi ^6$) of the ``bubble'' expansion
of the  action (\ref{3}).

Let us consider an  effective action (\ref{actrg}) with an
additional $F_3 \equiv \tr(\chi^{\dag}\chi)^3$ term.  In
renormalization procedure in $4-\varepsilon$ scheme $F_3$ will be
considered as a composite operator of canonical dimension $\Delta_3=
6-3\varepsilon$. Also, there are composite operators $F_2 \equiv
\tr(\chi^{\dag}\chi)^2 \tr(\chi^{\dag}\chi)$  and $F_1 \equiv
(\tr\chi^{\dag}\chi)^3$ with the same canonical dimension as $F_3$,
therefore they may be mixed in the process of  renormalization.
Thus the term $\lambda_{0j} F_j/36$ must be included in the
effective action, here $\lambda_{0j}$ are bare homogeneous sources.
One can define the set of renormalized parameters $\lambda _i$ using
$\lambda_{0j} = Z_{j k} \lambda_{k} M^{2\varepsilon-2}$ for such
extended  model, $Z_{jk}=\delta_{jk}+Z^1_{jk}/\varepsilon
+O(1/\varepsilon ^2)$. Matrix $Z$ is a function of the variables
$g_i$.  Similar to the (\ref{rg}) we can write RG functions for
$\lambda_j$

\begin{equation}
\label{rgl} \beta_{\lambda_j}=-(2\varepsilon-2)\lambda_j+\lambda_i
g_k\frac{\partial}{\partial g_k} Z^1_{ij}.
\end{equation}

The matrix $Z$ was calculated in the framework of three-loop
approximation, see appendix.  One-loop approximation of matrix $Z$
leads to the following results

\begin{equation}
\label{betal1} \beta_{\lambda_1}=2 (1 - \varepsilon) \lambda_1+ g_1
\left[{3 \over4} \lambda_1 ( r^2 - r + 14)+ \lambda_2 ( r - 1) + {3
\over 4} \lambda_3\right]+ {3 \over 2} g_2 \left[\lambda_1 ( r - 1)
+ \lambda_2 \right],
\end{equation}

\begin{equation}
\label{betal2} \beta_{\lambda_2}=2 (1 - \varepsilon)\lambda_2+ {1
\over 4} g_1 \left[ \lambda_3 (6 r - 9) + \lambda_2 ( r^2 - r + 38
)\right]+ {3 \over 2} g_2 \left[ 6 \lambda_1+ \lambda_2 ( r - 2) + 3
\lambda_3\right],
\end{equation}

\begin{equation}
\label{betal3} \beta_{\lambda_3}=2 (1 - \varepsilon) \lambda_3+{15
\over 2} g_1 \lambda_3+ {3 \over 2} g_2 \left[   \lambda_3 (r - 4) +
4 \lambda_2\right],
\end{equation}
the rescaling of charges $g_i \rightarrow g_i/{16\pi^2}$ is assumed.

Let us mark that the full family  of the composite operators with the
same canonical dimension in the $D=4$ dimensional space must be taken
into account for an accurate calculation of the $F_i$ operators renormalization.

This family also includes the operators
\begin{eqnarray}
\label{ff} f_2=\tr(\Delta \chi ^+ \Delta\chi), \quad
f_{41}=\tr(\Delta \chi^+\chi\chi^+ \chi)+ \tr(\chi^+\Delta\chi\chi^+
\chi),\quad f_{42}=\tr(\partial _i \chi^+\chi\partial _i\chi^+ \chi)
+\tr(\chi^+\partial _i\chi\chi^+ \partial _i\chi),\\
\nonumber
f_{43}=\tr(\Delta \chi^+\chi)\tr(\chi^+
\chi)+\tr(\chi^+\Delta\chi)\tr(\chi^+ \chi),\quad
f_{44}=\tr(\partial _i \chi^+\chi)\tr(\partial _i\chi^+
\chi)+\tr(\chi^+\partial _i\chi)\tr (\chi^+ \partial _i\chi),\quad
f_{45}=\tr(\partial _i \chi^+\partial _i\chi)\tr(\chi^+ \chi)
\end{eqnarray}
in addition to the $F_i$ operatorsd. The canonical
dimensions of these operators are $d[f_2]=D+2$, $d[f_{4i}]=2D-2$.
But in the analysis presented we limit ourselves by the
considerations of the $F_i$ operators only. The contributions of the
operators (\ref{ff}) will be discussed below.

It was shown in \cite{Nalimov} that  $\beta \equiv \left\langle \chi
\right\rangle$
 is an order parameter of phase transition in the model considered. A non-zero value of $\beta$ leads to phase transition to the superfluid phase.

 The  value for magnitude $\beta$ can be calculated by minimization of free energy $-\Gamma$.
  In the framework of the Landau mean field theory this functional can be written in the form

\begin{equation}
\label{lezh} -\Gamma =  \tau \tr\beta^{\dag}\beta
+\frac{g_{01}}{4} \left(\tr \beta\beta^{\dag}\right)^2 +
\frac{g_{02}}{4}\tr
\beta\beta^{\dag}\beta\beta^{\dag}+\frac{\lambda_{01}}{36} \left(\tr
\beta\beta^{\dag}\right)^3+\frac{\lambda_{02}}{36} \tr
\left(\beta\beta^{\dag}\right)^2\tr\beta\beta^{\dag}+\frac{\lambda_{03}}{36}\tr\left(\beta\beta^{\dag}\right)^3.
\end{equation}

\begin{figure}
\centering
\includegraphics[width=7cm]{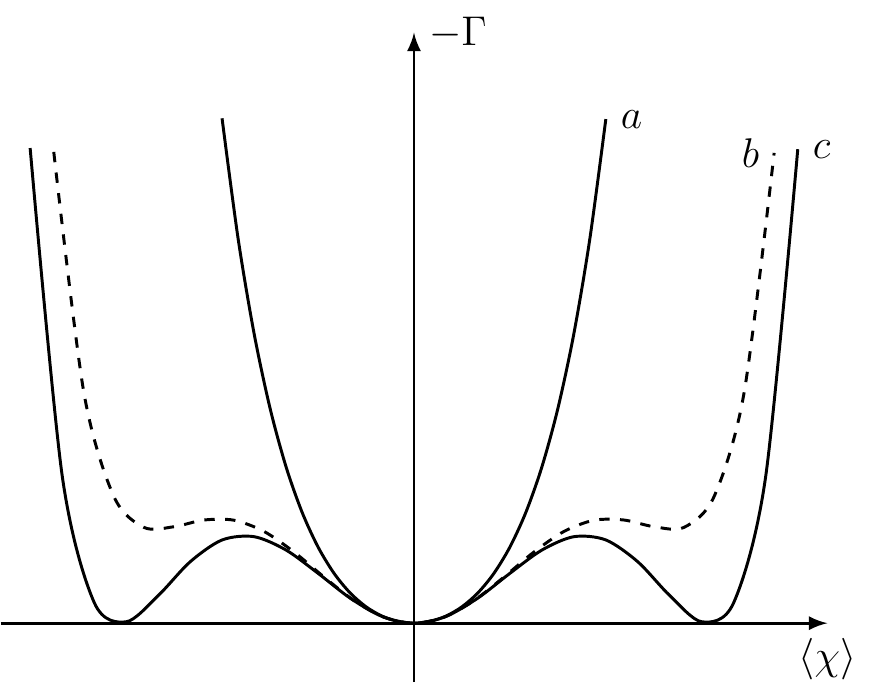}
\caption{Thermodynamics potential as a function of order parameter:
$a$ -- disorder state, $b$ -- metastable state, $c$ --
``superfluid'' state.} \label{risGam}
\end{figure}
 Schematically it can be represented in the figure (\ref{risGam}). The variables $\beta$ and $\beta^{\dag}$ have the Pfaffian's form
 (\ref{pff}). For the extrema conditions at the phase transition point
 we get
\begin{equation}
\label{exr} \frac{\partial}{\partial \beta_j}\Gamma=0,
\quad \frac{\partial}{\partial \beta^*_j}\Gamma=0, \quad
\Gamma=0 , \quad \forall j=0,\dots,r/2.
\end{equation}
Obviously, the loop corrections to equation \eqref{lezh} contains IR singularities. This singularities can be taken into account using RG method. This procedure leads to the
fact that charges $g_{0j}, \lambda_{0j}$ in (\ref{lezh}) must
now be replaced by the invariant charges  $\bar{g}_j,
\bar{\lambda}_j$, which in turn depend on the parameter $\tau$. After such processing, the contributions of higher
loops give only $\varepsilon$-corrections to the mean field theory
results. Let
us introduce $z_j\equiv \beta_j/M^{d_{\beta}}, s\equiv \tau/M^2$,
where $d_{\beta}=1-\varepsilon /2$ is canonical dimension of the field
$\beta$. Then RG equations for the invariant variables are

\begin{equation}
\label{gjrg}
\partial_{\xi} \bar{g}_j = \frac{\beta_{g_j}}{2+\gamma_{\tau}}, \quad \left.\bar{g}_j\right|_{\xi=0} = g_j,
\end{equation}
\begin{equation}
\label{ljrg}
\partial_{\xi} \bar{\lambda}_j = \frac{\beta_{\lambda_j}}{2+\gamma_{\tau}}, \quad \left.\bar{\lambda}_j\right|_{\xi=0} = \lambda_j,
\end{equation}
\begin{equation}
\label{betarg}
\partial_{\xi} \bar{z}_j = -\bar{z}_j \frac{\Delta_{\beta}+\gamma_{\beta}}{2+\gamma_{\tau}}, \quad \left.\bar{z}_j\right|_{\xi=0} = z_j,
\end{equation}
Finally, if we combine previous equations with condition (\ref{exr}) and
(\ref{lezh}), we get
\begin{equation}
\label{sol} |\bar{z}_j|^2 = -{9\over 2}\frac{2n
\bar{g}_1+\bar{g}_2}{4n^2\bar{\lambda}_1+2n\bar{\lambda}_2+\bar{\lambda}_3},
\end{equation}

\begin{equation}
\label{t} \tau= {9\over 16}\frac{(2n
\bar{g}_1+\bar{g}_2)^2}{4n^2\bar{\lambda}_1+2n\bar{\lambda}_2+\bar{\lambda}_3},
\end{equation}
 $n$ is number of non-zero blocks. Thus, as $\tau$ decreases, the invariant charges intersect the boundary of the stability domain and  new solution (\ref{sol}) of stationary equations (\ref{exr}) appears. This phase have two non-zero blocks, $n=2$. Equation (\ref{t}) determines the transition temperature $\tau_t$. In order to solve equation (\ref{t}) it is necessary to know solutions of RG equations (\ref{gjrg}) and (\ref{ljrg}). As before, the RG equations must be resummed. Similar to (\ref{g1}) we can rewrite the equations (\ref{ljrg})
 \begin{equation}
\begin{split}
\label{Lj}
&\partial_{\xi} {\bar \lambda}_i = -\frac{2\varepsilon-2}{\varepsilon}{\bar \lambda}_i+\sum\limits_{N=0}^{K} \varepsilon^N {\bar \lambda}_j L^{(N)}_{ji}({\bar g}_1,{\bar g}_2),\\
&\left.{\bar \lambda}_i\right|_{\xi=0} = \lambda_i.
\end{split}
\end{equation}
The HOA for $L^{(N)}_{ji}$ are needed  for the Borel resummation
too. Our analysis in Sect.2 shows that calculation of $L^{(N)}_{ji}$
coefficients is connected with the renormalization of six-point 1PI
Green functions ($\sim (\sqrt{N})^{6}$) which include one insertion
of composite operators $F_j \sim (\sqrt{N})^{6}$, hence
$L^{(N)}_{ji}/(\sqrt{N})^{6+6} \sim B^{(N)}_{i}/(\sqrt{N})^{4}$.
Indices structure is irrelevant for the HOA.

\begin{figure}
\centering
\includegraphics[width=7cm]{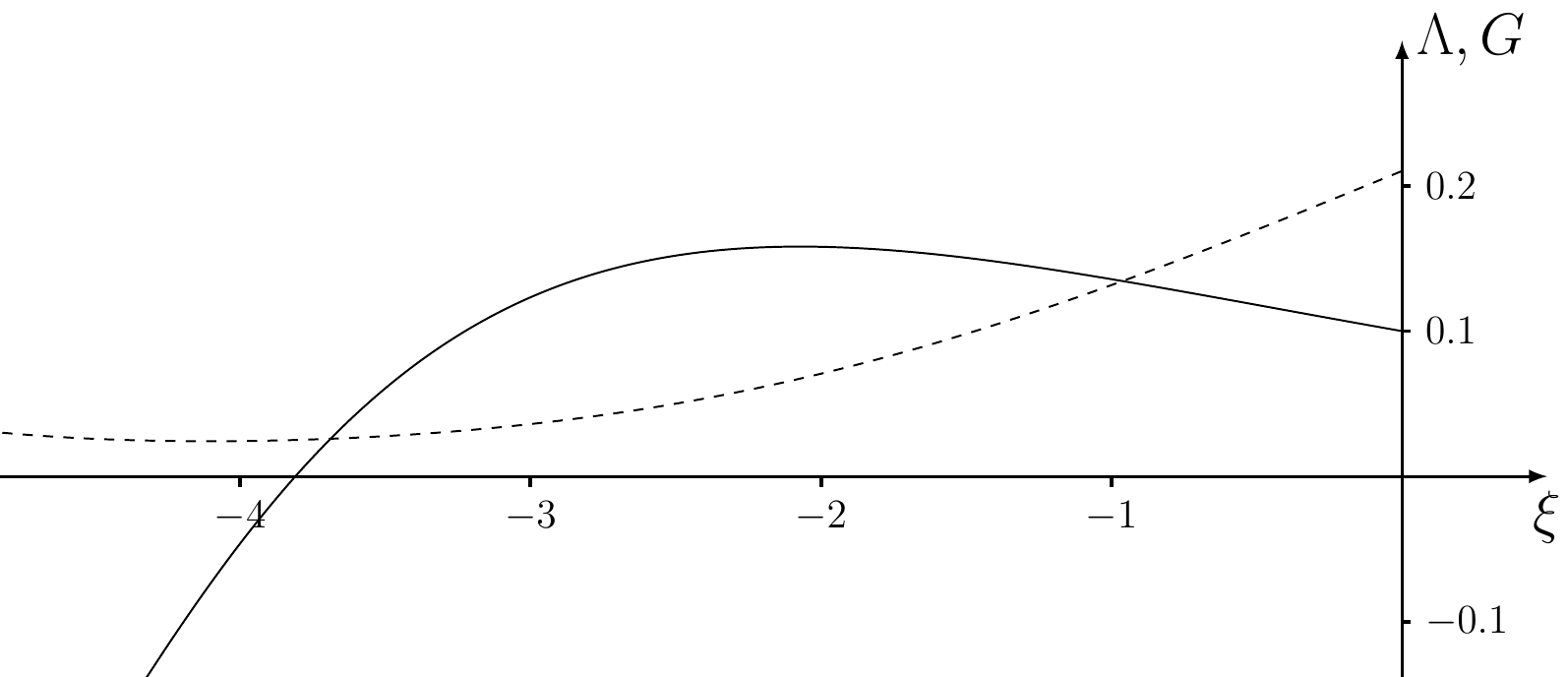}
\caption{Trajectories of ``effective'' coupling constants $\Lambda
\equiv 4 n^2 {\bar \lambda}_1 + 2 n {\bar \lambda}_2 + {\bar
\lambda}_3$ (dashed line) and $G \equiv 2 n {\bar g}_1 + {\bar g}_2$
(solid line) at $D=3$ and $n=2$.} \label{LG1}
\end{figure}

\begin{figure}
\centering
\includegraphics[width=7cm]{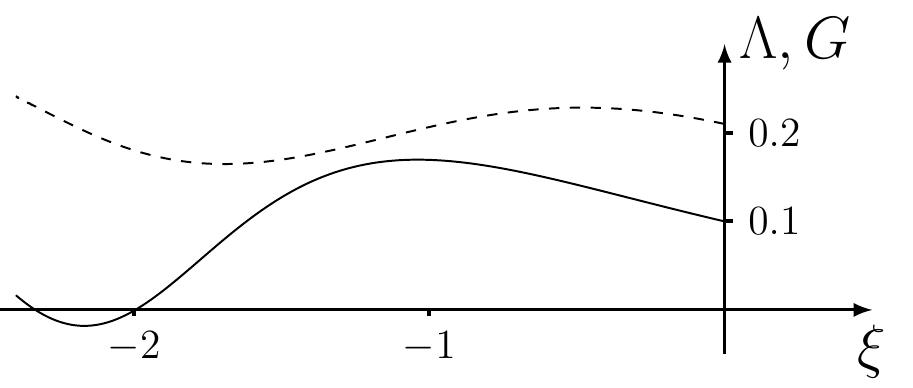}
\caption{Trajectories of ``effective'' coupling constants $\Lambda
\equiv  4 n^2 {\bar \lambda}_1 + 2 n {\bar \lambda}_2 + {\bar
\lambda}_3$ (dashed line) and $G \equiv 2 n {\bar g}_1 + {\bar g}_2$
(solid line) at $D=2$ and $n=2$.} \label{LG2}
\end{figure}

 Thus, we can resum the RG equations (\ref{Lj}) by the formula (\ref{cm_integ}).  The results of numerical computations are shown in figures
\ref{LG1} and \ref{LG2}. This allows us to solve the equation
(\ref{t}). As the result  the  root of this equation $\xi_c$ differs only
 a little from $\xi _0$ obtained in Sect.3. Remember that
$\xi _0$ demonstrates a weak dependence from initial values of the
coupling constants $g_i$.

Let us discus here a possible contributions of $f$ operators
(\ref{ff}) to the results obtained. There are some reasons why we
have not calculated these counterterms.

First, we can state that these contributions are relatively small
compared with $F$ because $f$ operators are more IR irrelevant in the
real space dimensions $D=2,3$ then $F$ according to the canonical
dimensions mentioned above. Then the corresponding invariant charges
will be oppressed by the first terms in the RG equations similar to
(\ref{Lj}) for these variables.

Second, it  can be simply shown by the instanton analysis presented
that the high-order contributions of $f$ operators are small in
$1/N$ compare with these of $F$.

And third, the calculations of the renormalization of the full
family of composite operators $F$ and $f$ is rather technically
difficult now. To calculate the full renormalization constants
matrix up to $\varepsilon^3$ corrections one needs to consider six-loop
diagrams. It is not worth while to start these calculations, as our results show that the first order phase transition takes place and the influence of $F$ operators on it's temperature is rather small.

Then we can state that in the model considered the first-order phase
transition takes place at a temperature higher than the predictions
for continuous phase transitions.

To estimate the temperature difference in $D=3$ case  we have calculated numerically that $\zeta_{0} = \tau_t/g_2^2 \approx 2 \div 3$  in a wide range of values $g_2 \approx 10^{-5} \div 0.1$, here $\tau_t$ is the root of the equation (\ref{t}). Natural to assume that the chargesare of the same order of magnitude $g_{1} \sim g_{2} << 1$. Then the renormalization constants $Z_d$ have the form  $Z_d = 1 + O(g_j)$.
In this approximation the ratio $Z_{\tau}/Z^2_{g_2}$ equals to $1$. This leads to the relation
\begin{equation}
\label{tauu}
\tau_0/g_{02}^2 = \zeta_{0}.
\end{equation}
The integrals over momenta and sum over frequencies $\omega_s$ (\ref{g}, \ref{c0}) can be reduced to the one-dimensional integrals. But the RG-approach  used in our article give us an opportunities to calculate different values for small $\tau$ only. Then it is sufficient  to calculate these parameters (\ref{g}, \ref{c0}) using the approximation $\beta \delta >> 1$. It can be found in this approximation
\begin{equation}
\label{gg} \quad \widetilde{g}_{02} \approx
\frac{7 \nu_F \beta}{8 (\pi T)^2}\zeta(3), \quad
\widetilde{\tau}_0\approx
{\beta\over 2\lambda} \left(1 - \lambda\nu_F \ln \frac{\gamma
\delta}{\pi T} \right), \quad
 c_0  \approx \frac{7 \nu_F p_F^2 \beta}{96 (\pi T m)^2}\zeta(3),
\end{equation}
with corrections $\sim O((\beta \delta)^{-1})$. Here $\nu_F = m p_F/(2 \pi^2)$ is $3D$-density of states at the Fermi level, $p_F$ is the Fermi momentum.

Near the transition point  $\tau_0$ can be estimated as
\begin{equation}
\label{c} \tau_{0} \approx \frac{\beta \nu_F}{2 c_0} \frac{\Delta T}{T_0},
\end{equation}
where $T_0$ is the continuous  phase transition temperature determent by the usual approach \cite{Abrikosov}. Combining (\ref{tauu}, \ref{gg}, \ref{c}) we get estimation for the temperature difference  between the first order phase transition and $T_0$
\begin{equation}
\label{deltat}
\frac{\Delta T}{T_0} = \zeta_0 \frac{6912 \pi^6}{7\zeta(3)}\left(\frac{T}{T_F}\right)^4.
\end{equation}

\section{Conclusions}

In contrast to the case of the electron systems ($r=2$, $r$ - number of spin degrees of freedom) where continuous phase transition takes place, our investigation has shown that in systems with high spin fermions ($r \geq 4$) critical fluctuations destroy stability of the system (see Fig.2). In such systems the first order phase transitions take place in space dimension $D=3$. These results were obtained by means of renormalization group analysis with $\varepsilon$-expansion up to the fifth-loop order of perturbation theory and subsequent Borel resummation. It should be noted that five loop calculations are indispensable to be sure that the first order phase transition takes place.

The temperature of transition to the superconducting or superfluid phase was estimated for the systems under consideration. Three loop RG analysis for composite operators, which are similar to $(\chi\chi^{\dag})^3$ in the Landau-Ginzburg functional, was performed for estimation of this temperature. It was revealed that transition temperature is higher than the theoretical estimation based on the continuous phase transition formalism for the same model. The obtained difference in temperatures is rather small (see expr. \eqref{deltat}).
But it should be kept in mind that approach used in the present work is applicable for the small deviations from the phase transition temperature only. Thus, in either case, we can guarantee that the difference in the phase transition temperature is not lower than our estimation.

As for 2D systems one can state that the five loop approximation is not sufficient to determine neither the phase transition type nor the phase transition temperature. The last is an excellent example in favor of further development of the high-loop calculations.

\section{Appendix}

The results obtained for coupling constants renormalization in
three-loop approximations are presented here. Five-loop results for
$\beta_{g_i}$, $\gamma_{\tau}$  and three-loop results for
$\beta_{\lambda_i}$ are available by kalagovg@gmail.com.
\begin{equation}
\label{b1}
\begin{split}
\beta_{g_1}& = -\varepsilon g_1+{1\over 4}(r^2-r+8)g_1^2+(r-1)g_1g_2+{3\over 4}g_2^2-\\
&-{9 \over 48}(3 r^2-3 r+14)g_1^3-{11\over 4}(r-1)g_1^2g_2-{1\over 32}(5 r^2-15 r+92)g_1g_2^2-{3 \over 8}(r-2)g_2^3+\\
&+{1\over 512}\left[33r^4-66r^3+(955+480\zeta(3))r^2-(480\zeta(3)+922)r+2960+2112\zeta(3)\right]g_1^4+\\
&+{1\over 128}\left[79 r^3-158 r^2+(1397+768\zeta(3))r-1318-768\zeta(3)\right]g_1^3g_2+\\
&+{1\over 1024}\left[3 r^4-12r^3+(576\zeta(3)+3355)r^2-(1728\zeta(3)+7568)r+9216\zeta(3)+14788\right]g_1^2g_2^2+\\
&+{1\over 512}\left[60 r^3-321 r^2+(2943+1152\zeta(3))r-2304\zeta(3)-4092\right]g_1g_2^3+\\
&+{1\over 1024}\left[(96 \zeta(3)+193)r^2-(576 \zeta(3)+891)r+1536
\zeta(3)+1860 \right]g_2^4,
\end{split}
\end{equation}
\begin{equation}
\label{b2}
\begin{split}
\beta_{g_2}&=-\varepsilon g_2+{1\over 4}(2r-5)g_2^2+3g_1g_2-\\
&-{3\over 32}(r^2-7r+20)g_2^3-{1\over 4}(11r-20)g_2^2g_1-{1\over 16}(5r^2-5r+82)g_1^2g_2+\\ &+{1\over 1024}\left[ 26r^3-(383+96\zeta(3))r^2+(2459+1152\zeta(3))r-4060-2688\zeta(3) \right]g_2^4+\\
&+{1\over 128}\left[(96\zeta(3)+182)r^2-(963+576\zeta(3))r+1536\zeta(3)+1937\right]g_2^3g_1+\\
&+{1\over 512}\left[-70r^3+11r^2+(6423+4608\zeta(3))r-8064\zeta(3)-10366\right]g_1^2g_2^2+\\
&+{1\over
256}\left[-13r^4+26r^3+(192\zeta(3)+355)r^2-(368+192\zeta(3))r+3284+2688\zeta(3)\right]g_1^3g_2
\end{split}
\end{equation}

\begin{equation}
\begin{split}
\label{gama}
\gamma_{\tau}&= -{1\over4}(r^2-r+2)g_1-{1\over 2}(r-1)g_2+{5\over 32}(r^2-r+2)g_1^2+{5\over8}(r-1)g_1 g_2+{5\over64}(r^2-3r+4)g_2^2\\
&-{1\over 256}(15 r^4-30 r^3+156 r^2-141 r+222)g_1^3-{3\over 128}(15 r^3-30 r^2+126r-111)g_1^2g_2\\
&-{3\over1024}(r^4-4r^3+403r^2-960r+888)g_1g_2^2-{1\over 1024}(90
r^3-321r^2+987r-888)g_2^3
\end{split}
\end{equation}

\section*{Acknowledgments.}

G.A. Kalagov and M.Yu. Nalimov are grateful to SPbSU grant 11.38.636.2013, M.V. Kompaniets is supported by SPbSU grant 11.38.185.2014.


\begin{thebibliography}{00}

\bibitem{Abrikosov}
A. A. Abrikosov, L. P. Gorkov, and I. E. Dzyaloshinski, {\it Methods
of Quantum Field Theory in Statistical Physics}, (Dobrosvet, Moscow
2006).

\bibitem{Vasil'ev1}
A.N. Vasil'ev, {\it Functional Methods in Quantum Field Theory and
Statistical Physics}, (Gordon and Breach, Amsterdam 1998).

\bibitem{Kagan}
M.A. Baranov, M.Yu. Kagan, Yu.Kagan, JETP Lett. 64:4 (1996) 273.


\bibitem{atoms}
R.W. Cherng, G. Refael, E. Demler, Phys. Rev. Lett. 99 (2007)
130406.

\bibitem{atoms1}
Tomoki Ozawa, Gordon Baym, Phys. Rev. A 82 (2010) 063615.

\bibitem{atoms2}
John L. Bohn, Phys. Rev. A  61 (2000) 053409.

\bibitem{atoms3}
Congjun Wu, Physics 3  (2010) 92.

\bibitem{atoms4}
Tin-Lun Ho, Sungkit Yip, Phys. Rev. Lett 82:2 (1999) 247.

\bibitem{atoms5}
Miguel A. Cazalilla, arXiv:1403.2792v1 (2014).

\bibitem{atoms6}
Masaru Sakaida, Norio Kawakami, Phys. Rev. A 90 (2014) 013632.


\bibitem{atoms7}
M. A. Cazalilla, A. F. Ho, M. Ueda, New Journal of Physics 11 (2009)
103033.

\bibitem{grafen}
M.I. Katsnelson {\it Graphene. Carbon in Two Dimensions}, (Camdridge
University press, 2012)

\bibitem{Gorkov}
L. P. Gor'kov, T. K. Melik-Barkhudarov, J. Exptl. Theoret. Phys. 40
(1961) 1452.

\bibitem{Nalimov}
J. Honkonen, M.V. Komarova, M.Yu. Nalimov, Theor. Math. Phys. 176:1
 (2013) 89.


\bibitem{Kalagov}
G. A. Kalagov, M. V. Kompaniets, M. Yu. Nalimov, Theor. Math. Phys.
181:2 (2014) 1448.

\bibitem{Vasil'ev}
A.N. Vasil'ev, {\it  Quantum-Field Renormalization Group in the
Critical Behavior Theory and in Stochastic Dynamics},  (St.
Petersburg Institute for Nuclear Physics, St. Petersburg, 1998).


\bibitem{Lipatov}
L. N. Lipatov, J. Exptl. Theoret. Phys. 72 (1977) 411.

\bibitem{Varm}
J.A.M.Vermaseren "New features of FORM" math-ph/0010025.

\bibitem{phi45loop}

Chetyrkin K. G., Kataev A. L. , Tkachev F. V., Phys.Lett. B 99 (1981) 147; Errata B 101 (1981) 457.\\
Chetyrkin K. G., Gorishny S. G., Larin S. A. and Tkachov F. V., Phys. Lett. B 132 (1983) 351. \\
Kazakov D. I. Phys.Lett. B 133:6 (1983) 406;  Theor.Math.Phys. 58 (1984)  223; Teor.Mat.Fiz. 58:3 (1984) 343.\\
Chetyrkin K. G.,  Gorishny S. G., Larin S. A. and Tkachov F. V., Preprint INR P-0453 (1986), Moscow.\\
Kleinert H., Neu J., Shulte-Frohlinde V., Chetyrkin K. G., Larin S. A., Phys.Lett. B 272 (1991) 39; Erratum B  319 (1993) 545.

\bibitem{phi45loopnum}
L.Ts. Adzhemyan, M.V. Kompaniets, Journal of Physics: Conference Series 523 (2014) 012049

\bibitem{HonkonenBH}
J. Honkonen, M.V. Komarova, M.Yu. Nalimov, Nuclear Physics B 714:3
(2005) 292.


\bibitem{Komarova}
M.V. Komarova, M. Yu. Nalimov, Theor. Math. Phys. 126:3 (2001) 339.

\bibitem{Zinn}
J. Zinn-Justin, {\it Quantum Field Theory and Critical Phenomena},
3rd edition (Clarendon Press, Oxford, 1996).

\bibitem{hagen}
Hagen Kleinert, Verena Schulte-Frohlinde, {\it Critical Properties
of $\phi^4$-Theories}, (World Scientific, Singapore, 2001).

\bibitem{sergeev}
M. Yu. Nalimov, V. A. Sergeev,L. Sladkoff, Theor. Math. Phys. 159:1
(2009) 96.
\end{thebibliography}
\end{document}